**THE MULTIVERSE**
**AND**
**COSMIC PROCREATION**
**VIA COFINSLER SPACES**
**OR**
**BEING AND NOTHINGNESS**

by

Gregory W. Horndeski
Adjunct Associate Professor of Applied Mathematics
University of Waterloo
Waterloo, Ontario
Canada
N2L 3G1

January 24, 2023

# ABSTRACT

In this paper I shall consider a scalar-scalar field theory with scalar field phi on a four-dimensional manifold M, and a Lorentzian Cofinsler function f on T*M. A particularly simple Lagrangian is chosen to govern this theory, and when f is chosen to generate FLRW metrics on M the Lagrangian becomes a function of phi and its first two time derivatives. The associated Hamiltonian is third-order and admits infinitely many vacuum solutions. These vacuum solutions can be pieced together to generate a multiverse. This is done for those FLRW spaces with k>0. So when time, t, is less than zero we have a universe in which the t=constant spaces are the 3-sphere with constant curvature k. As time passes through zero the underlying 4-space splits into an infinity of spaces (branches) with metric tensors that describe piecewise de Sitter spaces until some cutoff time, which will, in general, be different for different branches. After passing through the cutoff time all branches return to their original 4-space in which the t=constant spaces are of constant curvature k, but remain separate from all of the other branch universes. The metric tensor for this multiverse is everywhere continuous, but experiences discontinuous derivatives as the universe branches change between different de Sitter spaces. Questions I address using this formalism are: what is the nature of matter when t<0; what happens to matter as time passes through t=0; and what the universe was doing before the multiple universes came into existence at t=0. The answers presented will explain the paper's title. I shall also discuss a possible means of quantizing space, how inflation influences the basic cells that constitute space, and how gravitons might act.

**Section 1: Constructing a Multiverse**

To make this paper fairly self-contained I shall begin by reviewing some of the material on Lorentzian Cofinsler spaces and scalar-scalar field theories which we shall require. I presented these notions originally in Horndeski, Refs.1 and 2.

Let M be a 4-dimensional manifold with cotangent bundle T*M, and $\pi$:T*M→M the natural projection. If x is a chart of M with domain U then it naturally gives rise to a chart $(\chi,y)$ of T*M with domain $\pi^{-1}U$ defined by: $\chi := x \circ \pi$, and $\forall\ \omega \in T_p^*M$, with $p \in U$, $y(\omega) = y(\omega_i dx^i|_p) := (\omega_0,...,\omega_3) \equiv (\omega_i)$, where repeated indices are summed from 0 to 3. So if we write $y:\pi^{-1}U \to \mathbb{R}^4$ as $y=(y_0,...,y_3) \equiv (y_i)$, then $y_i(\omega) = \omega_i$. Now suppose that x and x' are overlapping charts of M with domains U and U'. On the overlap $\pi^{-1}(U \cap U')$ we have

$$\chi^i = \chi^i(\chi^{ij})\ ,\ \chi^{ti} = \chi^{ti}(\chi^j) \qquad\qquad\text{Eq.1.1}$$

and

$$y_i = y'_j\frac{\partial x^{tj}}{\partial x^i} \circ \pi = y'_j\frac{\partial \chi^{tj}}{\partial \chi^i}\ \ ,\ y'_i = y_j\frac{\partial x^j}{\partial x^{ti}} \circ \pi = y_j\frac{\partial \chi^j}{\partial \chi^{ti}} \qquad\text{Eq.1.2}$$

I define a 4-dimensional Lorentzian Cofinsler space as follows. Let f:T*M→$\mathbb{R}$ be a smooth function defined on an open submanifold N of T*M, where $\pi$(N) = M. If x is a chart of M with domain U and corresponding standard chart $(\chi,y)$ of T*M with domain $\pi^{-1}U$, then $\forall\ \omega \in \pi^{-1}U \cap N$, I require that the matrix

$$\left[\frac{\frac{1}{2}\partial^2 f}{\partial y_i \partial y_j}\right]\bigg|_\omega \qquad\qquad\text{Eq.1.3}$$

defines a Lorentzian quadratic form on $\mathbb{R}^4$, with signature (−+++). Due to Eqs.1.1 and 1.2 this restriction on f is independent of the chart x. When f satisfies these conditions the triple $CF^4 :=$ (M,N,f) is called a 4-dimensional Lorentzian Cofinsler space and f is called a Lorentzian Cofinsler function.

Suppose that we have a 4-dimensional Lorentzian Cofinsler space $CF^4 =$ (M,N,f) and a scalar field φ on M.  If dφ:M→T*M has its range contained in N then we can define a Lorentzian metric tensor $g_φ$ on M by stipulating that its contravariant components on the domain of any chart x of M are

$$g_φ{}^{ij} := g_φ(dx^i, dx^j) := \tfrac{1}{2} \frac{\partial^2 f}{\partial y_i \partial y_j}(dφ) \ .$$
<div align="right">Eq.1.4</div>

I call a Lorentzian Space $V_4 = (M,g)$ endowed with a scalar field φ in which the metric tensor g arises from a Lorentzian Cofinsler space $CF^4 = (M,N,f)$ in the manner presented in Eq.1.4 a scalar-scalar theory.  So formally a scalar-scalar theory is a pair $(CF^4=(M,N,f), φ)$ where $dφ(M) \subset N$.  In such theories the two scalar functions f and φ act as generating functions for the metric tensor.

If one is given an 4-dimensional Lorentzian space, $V_4 = (M,g)$, it is always possible to build a Lorentzian Cofinsler space $CF^4 = (M,N,f_T)$ such that the metric $f_T$ generates, with any scalar field φ on M, is the g of your $V_4$.  For just locally define the Cofinser function $f_T$ by

$$f_T := (g^{ij} \circ \pi) \, y_i y_j \ .$$
<div align="right">Eq.1.5</div>

This Cofinsler function, and corresponding $CF^4$ are said to be trivial.  You should note that $f_T$ is homogeneous of degree 2 in the $y_i$'s; *i.e.,* $f_T(\lambda \omega) = \lambda^2 f_T(\omega)$, $\forall \ \omega \in N$, and $\lambda \in \mathbb{R}$.  In general this is not true for Cofinsler functions. It also explains why if you start with a scalar-scalar theory, and then build $f_T$ using Eq.1.5 with $g^{ij}$ replaced by the $g_φ{}^{ij}$ of your theory, you will not recover your original f. In general f contains more information than $f_T$.

In this paper we shall need a scalar-scalar theory whose associated metric is similar to the FLRW (:=Friedmann, Ref.3; Lemaître, Ref. 4; Robertson, Ref.5; Walker, Ref.6) metric. In reduced circumference polar coordinates the FLRW metric is given by

$$ds^2 = -dt^2 + a(t)^2((1-kr^2)^{-1}dr^2 + r^2 d\theta^2 + r^2\sin^2\theta d\zeta^2) \ ,$$
<div align="right">Eq.1.6</div>

where k is a constant with units of $(length)^{-2}$, r has units of length, a(t) is unitless, and c=G=1, with θ and ζ denoting the usual polar coordinates on the 2-sphere, $S^2$.  k determines the curvature of the 3-spaces t = constant, which are surfaces of constant curvature.

For the present I shall restrict our attention to the case k>0.  (In Ref.2, the k=0 case is treated in detail, and I shall remark about that case again in Section 4.)  The underling differentiable manifold of our spacetime will be M := $\mathbb{R} \times S^3$, where $S^3$ is the 3-sphere. I shall let $S_k$ denote the 3-dimensional Riemannian manifold with underlying manifold $S^3$ and line element

$$d\Omega^2 := (1-kr^2)^{-1}dr^2 + r^2d\theta^2 + r^2\sin^2\theta d\zeta^2 \qquad \text{Eq.1.7}$$

For later use I let $V_k$ be the 4-dimensional Lorentzian manifold with underling space $\mathbb{R} \times S^3$ and line element $ds^2 := -dt^2 + d\Omega^2$.  $V_k^-$ will denote the open submanifold  of $V_k$ with t<0.

If $(t,r,\theta,\zeta)$ denotes the aforementioned local coordinates in M then $((t\circ\pi,\ r\circ\pi,\ \theta\circ\pi,\ \zeta\circ\pi),\ (y_t,\ y_r,\ y_\theta,\ y_\zeta))$ denotes local coordinates of the associated standard chart of T*M.  Let f:T*M→$\mathbb{R}$ be defined on the domain of a standard chart by

$$f:= -y_t^2 + y_t^{-2}((1-k(r\circ\pi)^2)y_r^2 + (r\circ\pi)^{-2}y_\theta^2 + (r\circ\pi)^{-2}\sin^{-2}(\theta\circ\pi)\ y_\zeta^2)\ . \qquad \text{Eq.1.8}$$

f is smooth at all points of this chart where $y_t \neq 0$. It is a straightforward matter to demonstrate that f defines a Lorentzian Cofinsler function at all points of this chart where $y_t \neq 0$.  Due to the product nature of both M=$\mathbb{R} \times S^3$, and the charts that we are employing, we can identify  T*M  with T*$\mathbb{R} \times$T*$S^3$.  Thus if we define N:= [T*$\mathbb{R} \setminus$\{zero section\}]$\times$T*$S^3$, the function f defined by Eq.1.8 will define a Lorentzian Cofinsler function on N.

If $\varphi=\varphi(t)$ is a scalar field on M, then we can use Eqs.1.4 and 1.8 to show

$$[g_\varphi^{ij}] = \left[\tfrac{1}{2}\frac{\partial^2 f}{\partial y_i \partial y_j}\right](d\varphi) \quad = \text{diag}(-1,\ \varphi'^{-2}(1-kr^2),\ \varphi'^{-2}r^{-2},\ \varphi'^{-2}r^{-2}\sin^{-2}\theta)\ , \qquad \text{Eq.1.9}$$

where a prime denotes a derivative with respect to t.  The metric tensor presented in Eq.1.9 corresponds to the k>0, FLRW metric given in Eq.1.6 when we set a(t) = $\varphi'$(t).  Thus the scalar-scalar theory given by ((M,N,f),$\varphi$ ) with f as in Eq.1.8 does give rise to a FLRW spacetime.  In what follows I shall drop the subscript $\varphi$ on the metric tensor $g_\varphi$ given in Eq.1.9.

Earlier I mentioned units relevant to the FLRW metric. I would now like to say a few more things about the units we shall use. Since I have already selected our units so that c=G=1, we can measure all dimensioned quantites in terms of length, $\ell$. To indicate that a physical quantity Q has units of length to the power $\lambda$, $\ell^\lambda$, I shall write $Q\sim\ell^\lambda$. The units for our various coordinates for T*M will be chosen so that $ds^2\sim\ell^2$, $f\sim\ell^0$, $\varphi\sim\ell^1$, and $\int Ldtdrd\theta d\zeta\sim\ell^2$, so $L\sim\ell^0$.

In Ref.2 I argue that a suitable Lagrange scalar density for our k>0, FLRW scalar-scalar theory is provided by

$$L_f := -\tfrac{1}{8}\kappa g^{\frac{1}{2}}\varphi(f^*)^{-4}g^{ab}(f^*)_{,a}(f^*)_{,b} \qquad\qquad \text{Eq.1.10}$$

where $g:=|\det(g_{ab})|$, $f^*:=d\varphi^*f\equiv f\circ d\varphi$, and $\kappa\sim\ell^{-1}$ is a constant. Since we desire f to be given by Eq.1.8 we can use Eq.1.10 write

$$L_f = \tfrac{1}{2}\kappa[(1-kr^2)^{-\frac{1}{2}}r^2\sin\theta]\varphi(\varphi')^{-3}(\varphi'')^2 , \qquad\qquad \text{Eq.1.11}$$

where I have assumed that $\varphi'>0$, so that $g^{1/2}=(\varphi')^3(1-kr^2)^{-\frac{1}{2}}r^2\sin\theta$. Note that since $\varphi$ is independent of r, $\theta$ and $\zeta$ we can really treat $L_f$ as if it were only a function of t from the point of view of the calculus of variations.

I realize that $L_f$ is a nondegenerate second-order Lagrangian and as such is usually shunned because it can lead to multiple vacuum state solutions and other problems (*see*, Woodard, Ref.7). But I feel that multiple vacuum states are precisely what we are looking for in our quest for a multiverse.

For a Lagrangain which is a function of $\varphi$, $\varphi'$ and $\varphi''$ the Ostrogradsky Hamiltonian is given by (*see*, Woodard, Ref. 7, Ostrogradsky, Ref.8)

$$H := P_1\varphi' + P_2\varphi'' - L, \qquad\qquad \text{Eq.1.12}$$

where

$$P_1 := \frac{\partial L}{\partial\varphi'} - \frac{d}{dt}\frac{\partial L}{\partial\varphi''} \quad\text{and}\quad P_2 := \frac{\partial L}{\partial\varphi''} \qquad . \qquad\qquad \text{Eq.1.13}$$

One useful property of the Ostrogradsky Hamiltonian is that

$$\underline{\frac{dH}{dt}} = -\varphi' \underline{\frac{\delta L}{\delta \varphi}} \quad , \qquad\qquad\qquad\qquad \text{Eq.1.14}$$

where the term on the right-hand side of Eq.1.14 multiplying $\varphi'$ is the variational derivative of L.

Consequently the equation H= "constant" (strictly speaking H = function of r, $\theta$, $\zeta$) is a first integral

of the field equations. We shall be interested in the solutions to H = 0, which I shall call the vacuum

solutions, even though when L is chosen to be $L_f$ the corresponding Hamiltonian $H_f$ has solutions to

$H_f$ = constant<0.

Using Eqs.1.11-1.13 we easily find that

$$H_f = -\kappa[(1-kr^2)^{-\frac{1}{2}}r^2\sin\theta]\underline{\frac{d}{dt}}[\varphi(\varphi')^{-2}\varphi'']. \qquad\qquad\qquad \text{Eq.1.15}$$

All of the solutions to $H_f$ = 0 are given by

$$\varphi = \alpha e^{\beta t} \text{ and } \varphi = \gamma(\varepsilon_1 t + \varepsilon_2)^q \qquad\qquad\qquad\qquad \text{Eq.1.16}$$

where $\alpha \sim \ell^1$, $\beta \sim \ell^{-1}$, $\gamma \sim \ell^1$, $\varepsilon_1 \sim \ell^{-1}$, $\varepsilon_2 \sim \ell^0$ and $q \sim \ell^0$ are real numbers chosen so that $\varphi$ is well-defined with

$\varphi'$>0. The exponential solutions provide us with de Sitter geometries, Ref.9. We shall only be

interested in the second class of solutions when q = 1, in which case $\gamma$ can be absorbed into $\varepsilon_1$ and

$\varepsilon_2$, thereby changing their units. Thus we see that the scalar-scalar theory governed by the

Lagrangian $L_f$ provides us with infinitely many vacuum solutions, and that is the good news.

In passing I would like to point out that if we were looking for solutions to $H_f$ = "constant" $\neq$

0, then we can choose that "constant" to have the form $-\kappa[(1-kr^2)^{-\frac{1}{2}}r^2\sin\theta]\omega$, where $\omega$ is a constant.

Hence the equation $H_f$ = "constant" becomes

$$\underline{\frac{d}{dt}}[\varphi(\varphi')^{-2}\varphi''] = \omega \quad .$$

I shall now present an example to demonstrate how the various vacuum solutions presented

in Eq.1.16 can be pieced together to get a universe similar to the traditional inflationary universe of

Guth, Ref.10. To that end consider the scalar field defined by

$$\varphi := \begin{cases} \varphi_0 = t, \text{ for } t<0, \\ \varphi_1 = \beta_1^{-1}\exp(\beta_1 t), \ 0 \leq t < t_1 \quad \text{and} \\ \varphi_2 = \alpha_2\exp(\beta_2 t) \ , \ t_1 \leq t \end{cases} \qquad \text{Eq.1.17}$$

where $t_1$, $\beta_1$ and $\alpha_2$ are chosen so that $\varphi_1'(t_1) = \varphi_2'(t_1)$, $\beta_1 > 0$ and $\alpha_2\beta_2 > 0$ (so that $\varphi' > 0$). The function $\varphi$ is clearly going to have discontinuities at $t=0$ and $t=t_1$, although $\varphi'$ will be continuous at these points if we use left and right hand derivatives, which is what we shall do throughout this paper. So we can use Eqs.1.9 and 1.17 to construct a universe which begins as the 4-dimensional Lorentzian space $V_k^-$ introduced after Eq.1.7, with the line element

$$ds^2 = -dt^2 + (1-kr^2)^{-1}dr^2 + r^2 d\theta^2 + r^2\sin^2\theta \ d\zeta^2 \ .$$

At $t=0$, where $\varphi'(0)=1$, this universe begins to expand exponentially, and then at time $t=t_1$ it jumps to another exponentially expanding (or contracting, if $\beta_2 < 0$) universe in such a way that the metric tensor is continuous for all time. Moreover, for this universe the field equations generated by $L_f$ are satisfied for all times except $t=0$ and $t=t_1$, where the metric is not differentiable. This universe, would be my version of a universe suitable for Guth's original inflationary universe.

It is significant that this construction yields a continuous metric tensor because that guarantees the integrity of the spacetime causal structure. If $g$ were discontinuous at $t=0$ or $t=t_1$ then we could have timelike worldlines becoming spacelike, and vice versa, which would create chaos.

For the general FLRW metric given in Eq.1.6 it can be shown that when $a(t) = \varphi'(t)$,

$$R = 6(\varphi')^{-2}[\varphi'\varphi''' + (\varphi'')^2 + k]$$

and

$$R^{abcd}R_{abcd} = 12[(\varphi'''/\varphi')^2 + (\varphi')^{-4}[(\varphi'')^2 + k]^2] \ .$$

Thus for the scalar field presented in Eq.1.17:

(i) The scalar curvature R jumps from $R=6k$ to $R=12\beta_1^2 + 6k$, as we pass from $\varphi_0$ to $\varphi_1$ at $t=0$; and

(ii) R jumps from $12\beta_1^2 + 6k\exp(-2\beta_1 t_1)$ to $12\beta_2^2 + 6k\exp(-2\beta_1 t_1)$ as we pass from $\varphi_1$ to $\varphi_2$ at $t=t_1$, where I have used the fact that $\varphi_1'(t_1) = \varphi_2'(t_1)$.

In Guth's inflationary models $\beta_1$ is quite large and so unsuprisingly the jump in R as time passes through t=0 would be very large in that case. Also in Guth's models $\beta_2$ is very close to $0^+$, in which case there would be a shocking reduction in R as time passes through t=t$_1$. In the context of the present theory you would think that drastic reduction of R would leave its mark on the universe.

I shall now add a fourth branch to the scalar field given in Eq.1.17, that will return the universe described so far back to the "initial space" $V_k$ at time $t_c>t_1$. To that end let

$$\varphi_3 := \alpha_3 \exp(\beta_3 t),\ t_2 \leq t < t_c \qquad\qquad \text{Eq.1.18}$$

where I assume that $t_c>t_2>t_1>0$ have been chosen so that

$$\varphi_2'(t_2) = \varphi_3'(t_2) > 1,\ \text{and}\ \ \varphi_3'(t_c) = 1\ , \qquad\qquad \text{Eq.1.19}$$

and the domain of $\varphi_2$ is now $t_1 \leq t < t_2$. The reason I require $\varphi_3'(t_2)>1$, is to guarantee that this fourth branch of the universe does not start out spatially smaller than $S_k$. If $\varphi_3'(t_2)<1$, then this branch would have to expand exponentially to reach $V_k$. I prefer to think of $V_k$ as the "ground state" of the universe from which exponentially expanding sections of the universe can "bubble up," and then exponentially shrink back into $V_k$.

So what I have devised here is a universe that starts out as $V_k$, a static four-dimensionally Lorentzian space with t=constant slices of constant curvature k, and then at t=0 expands exponentially for a while. This initial expansion is then followed by two other exponential phases which eventually bring us back to the point where the scale factor $\varphi'$ reduces to 1 at time $t_c$ (crunch time). At this time the universe could return to the space $V_k$ with $t>t_c$, or begin expanding again with $\varphi$ having the form of $\varphi_1$ in Eq.1.17. This expanding and contracting universe could continue forever with the metric always being continuous and piecewise differentiable. And now that you see how this game is played, you probably realize that you did not need to begin in $V_k$ when t<0. The universe could "begin" at t=0 as the the tail end of a contracting $\varphi_3$ phase, with $t_c$ for that phase being t=0. This "bouncing" or "breathing" universe, that expands and contracts for all time will have a continuous metric tensor that satisfies the fields equations except at the discrete times where $\varphi$ is discontinuous. This bouncing universe is different than the cyclic universe of P. J. Steinhardt and N. Turok Ref.11, and I. Bars, *et al.* Ref 12.

In what follows I shall refer to a real valued function $\psi=\psi(t)$ which is piecewise of Class $C^k$ ($k\geq1$) but has discrete discontinuities at which the left and right hand first derivatives exist and are equal, as being of class $C^{k,1}$. By the left and right hand first derivatives of $\psi$ at a point $\tau$ of discontinuity for $\psi$ I mean $\lim\limits_{t\to\tau^-} \psi'(t)$ and $\lim\limits_{t\to\tau^+} \psi'(t)$ respectively.

I shall now explain how we can go about constructing a multiverse. To that end let

$$\mathbf{P_5}:= \{(\beta_1,\alpha_2,\beta_2,\alpha_3,\beta_3) \in \mathbb{R}^+\times\mathbb{R}^4 | \exists\ t_c>t_2> t_1>0, \text{ with } \exp(\beta_1 t_1)=\alpha_2\beta_2\exp(\beta_2 t_1),\ \alpha_2\beta_2\exp(\beta_2 t_2)=$$

$$= \alpha_3\beta_3\exp(\beta_3 t_2)>1 \text{ and } \alpha_3\beta_3\exp(\beta_3 t_c)=1, \text{ with } \beta_i\neq\beta_{i+1}, i=1,2\} \ . \qquad \text{Eq.1.20}$$

For each $(\beta_1,\alpha_2,\beta_2,\alpha_3,\beta_3)\in\mathbf{P_5}\subset\mathbb{R}^5$ we can build a universe from $\varphi$ constructed using Eqs.1.17 and 1.18. Each of these universes begins as a $V_k^-$ with the underlying manifold being $\mathbb{R}^-\times S^3$. I shall call these t<0 open submanifolds of each of these universes the universe's "tail." We could take the disjoint union of all these universes as $(\beta_1,\alpha_2,\beta_2,\alpha_3,\beta_3)$ ranges over $\mathbf{P_5}$ and then define an equivalence relation on that union which essentially enables us to glue all of these tails together. The result would give rise to a multiverse as a quotient manifold (*see,* Brickell & Clark, Ref. 13 for the manifold theory used here) in which all of the $\varphi$s (and corresponding metric tensors) on the "leaves" of the disjoint union would project to a single function $\varphi$ (and metric tensor) on the quotient space. However, there is a second approach to constructing the multiverse which I shall describe next. I believe that this second approach is more informative since it is based on the idea of multifurcating the usual 1-dimensional time line at the point t=0. In fact, one could actually use this second approach to put the manifold structure on the quotient manifold described above, and so these two approaches lead to the same result, up to diffeomorphism.

Let $P_n$ be a subset of $\mathbb{R}^n$. In our applications $P_n$ will comprise the parameters which characterize different universes in a multiverse. These different universes all have a global time function since they will be of the FLRW type. In addition all of these universes will emanate from the Lorentzian space $V_k^-$, which is an open submanifold of $V_k$, with underlying manifold $\mathbb{R}^- \times S^3$. So we want the global time function on the $\mathbb{R}$ part of $V_k$ to split; *i.e.,* multifurcate as time passes through t=0, to obtain an immense 1-dimensional manifold that will replace the $\mathbb{R}$ in $\mathbb{R} \times S^3$. This 1-dimensional manifold will reside in $\mathbb{R} \times \mathbb{R}^n$ and is built as follows. (The ideas behind my construction of this manifold came from Problem 3.2.1 on page 40 of Ref.13.)

$\forall \, p = (p_1,...,p_n) \in P_n$ let $U_p(P_n) \subset \mathbb{R}^{n+1}$ be defined by

$U_p(P_n) := \{(t,0,...,0) \in \mathbb{R}^{n+1} | t<0\} \cup \{(t,p_1,...,p_n) | t \geq 0\}$ ,

and let $\boldsymbol{T}(P_n) := \bigcup_{p \in P_n} U_p(P_n)$ . I define a 1-dimensional chart $t_p$ for $\boldsymbol{T}(P_n)$ with domain $U_p(P_n)$ by

$t_p((t,0,...,0)) := t$ and $t_p((t,p_1,...,p_n)) := t$ .

It is clear that if $p,p' \in P_n$ then the two charts $t_p$ and $t_{p'}$ are $C^\infty$ related. The collection of charts $C(\boldsymbol{T}(P_n)) := \{t_p | p \in P_n\}$ determines a $C^\infty$ structure on dimension 1 on $\boldsymbol{T}(P_n)$. With this structure $\boldsymbol{T}(P_n)$ is a differentiable manifold of dimension 1. In terms of the manifold topology, which is determined by the complete atlas associated with $C(\boldsymbol{T}(P_n))$, $\boldsymbol{T}(P_n)$ is connected and satisfies the $T_1$ separation axiom. But it is not Hausdorff since it is impossible to separate the points (0,p) and (0,p') when $p \neq p'$. $\boldsymbol{T}(P_n)$ has a countable basis for its topology if and only if $P_n$ is a countable subset of $\mathbb{R}^n$ (see Proposition 3.3.3 in Ref.13).

I define a global time function t on $\boldsymbol{T}(P_n)$ by $t := t_p$ on the domain of each chart $t_p$. So we see that time "fractures;" *i.e.,* multifurcates, as we pass through $t = 0^-$. I shall refer to coordinate domains $U_p(P_n)$ as branches of $\boldsymbol{T}(P_n)$. These branches all merge when $t < 0$.

The time manifold $\boldsymbol{T}(P_n)$ can be used to build a multiverse, $\mathbf{MV_k(P_5)}$, as follows. Let

$$\mathbf{MV_k(P_5)} := \boldsymbol{T}(\mathbf{P_5}) \times S^3 \ . \qquad\qquad \text{Eq.1.21}$$

A geometric structure consistent with our solutions to $H_f = 0$ can be defined on $\mathbf{MV_k(P_5)}$ in the following manner. Let $p = (\beta_1, \alpha_2, \beta_2, \alpha_3, \beta_3) \in \mathbf{P_5}$, and let $t_p$ be the corresponding chart of $\boldsymbol{T}(\mathbf{P_5})$ with domain $U_p(\mathbf{P_5})$. If $t$ denotes the global time function on $\boldsymbol{T}(\mathbf{P_5})$ then on the branch $U_p(\mathbf{P_5}) \times S^3$ of $\mathbf{MV_k(P_5)}$ I set

$$\varphi := \begin{cases} t, \text{ if } t < 0, \\ \beta_1{}^{-1}\exp(\beta_1 t), \text{ if } 0 \le t < t_1 \\ \alpha_2 \exp(\beta_2 t), \quad \text{if } t_1 \le t < t_2 \\ \alpha_3 \exp(\beta_3 t), \quad \text{if } t_2 \le t < t_c, \\ t, \text{ if } t_c \le t \ . \end{cases} \qquad \text{Eq.1.22}$$

$\varphi$ is evidently well-defined on $\mathbf{MV_k(P_5)}$ since if $p$, $p' \in \mathbf{P_5}$ then on the intersection of their corresponding branches $U_p(\mathbf{P_5}) \times S^3$ and $U_{p'}(\mathbf{P_5}) \times S^3$, $\varphi = t$. Thus $\varphi$ is of class $C^{\infty,1}$ on each branch of $\mathbf{MV_k(P_5)}$ and satisfies $H_f = 0$, except where $\varphi$ is discontinuous. Apparently the Lorentzian Cofinsler functions f defined on each branch of $\mathbf{T^*MV_k(P_5)}$ by Eq.1.8, can be combined to yield a Lorentzian Cofinsler function on all of $\mathbf{T^*MV_k(P_5)}$, which I shall also denote by f. If we combine this f with the function $\varphi$ given in Eq.1.22, then Eq.1.4 permits us to construct a metric tensor on $\mathbf{MV_k(P_5)}$ which agrees with the metric tensors already present on each branch of $\mathbf{MV_k(P_5)}$. This metric tensor will be continuous and piecewise of class $C^\infty$.

The parameter space $\mathbf{P_5}$ corresponds to a universe with three exponential phases. We can also define a parameter space $\mathbf{P_{1+2n}} \subset \mathbb{R}^+ \times \mathbb{R}^{2n}$ for $n > 2$, in which the associated universe would have $n+1$ exponential phases beginning in a $V_k{}^-$ and ending in a $V_k$ with $t > t_c > 0$. In Ref.2, I show that for all of these universes the value of the action integral obtained by integrating $L_f$ over the entire universe will be zero. At first one might think that is great, all of our various solutions minimize the value of the action. But in the course of doing that analysis one sees that you could have a universe built from a $P_s$ parameter, and then you could add to that universe portions that expand and contract exponentially during (say) the second exponential phase without affecting the action. Such a result seems kind of bizarre, since we do not see such behavior in our universe. I would hope that such wildly fluctuating vacuum universes would not lead to minimizing actions. In Ref.2, I demonstrate that it is possible to introduce a second scalar field $\xi \sim \ell$, on $MV_k(P_s)$ and modify our initial Lagrangian $L_f$ to obtain a more well behave theory from a Lagrangian $L_T$ defined by

$$L_T := L_f + L_\xi \qquad\qquad \text{Eq.1.23}$$

where $L_f$ is defined by Eq.1.10 and

$$L_\xi := -\tau g^{\frac{1}{2}} \varphi^2 |f^*|^{-5/2}\, g^{ij}\xi_i\xi_j \qquad\qquad \text{Eq.1.24}$$

with $\tau \sim \ell^{-4}$ a constant. For the Lorentzian Cofinsler function we have been working with, Eqs. 1.11, 1.23 and 1.24 can be combined to show that when $\xi = \xi(t)$

$$L_T = ((1-kr^2)^{-\frac{1}{2}} r^2 \sin\theta)[\tfrac{1}{2}\kappa\varphi(\varphi')^{-3}(\varphi'')^2 + \tau\varphi^2(\varphi')^{-2}(\xi')^2]\ . \qquad\qquad \text{Eq.1.25}$$

In Ref.2, I show that all of the solutions to the Euler-Lagrange equations associated with $L_T$ beginning from a $V_k^-$ are given by

$$\varphi = \alpha e^{\beta t},\ \ \xi = \mu\beta^2 t + \xi_0 \ \ \text{and}\ \ \varphi = t,\ \xi = \mu t^{-1} + \xi_0 \qquad\qquad \text{Eq.1.26}$$

where $\mu \sim \ell^2$ and $\xi_0 \sim \ell$ are constants. (In Ref.2 it is shown that $\mu$ depends upon the branch of $\varphi$ that you are on. For the $\varphi = t$ solution $\mu$ must be zero for $t<0$, and so I also take $\mu=0$, when $\varphi=t$, $t>0$, since it does not seem reasonable to have $\xi$ non-constant when the geometry of space is stationary.) Hence we can still build our universe models that begin and end in a $V_k$ for $t<0$ and $t>t_c>0$, with de Sitter spaces linked together between these initial and final states. For these models the scalar field $\xi$ is characterized by one constant, $\mu>0$. So for our two scalar field $\varphi$, $\xi$ models the parameter space

$\mathbf{P_5}$ is replace by

$$\mathbf{P_{5,1}} := \{((\beta_1,\alpha_2,\beta_2,\alpha_3,\beta_3),\mu)\in\mathbf{P_5}\times\mathbb{R}^+\} \qquad\qquad \text{Eq.1.27}$$

where the $\xi$ solution corresponding to $(\beta_1,\alpha_2,\beta_2,\alpha_3,\beta_3)\in\mathbf{P_5}$ is

$$\xi := \begin{cases} 0, \text{ if } t<0, \\ \mu\beta_1{}^2 t, \ 0{\leq}t<t_1 \\ \mu\beta_2{}^2(t-t_1)+\mu\beta_1{}^2 t_1, \ t_1{\leq}t<t_2 \\ \mu\beta_3{}^2(t-t_2)+\mu\beta_2{}^2(t_2-t_1)+\mu\beta_1{}^2 t_1, \ t_2{\leq}t<t_c \\ \mu\beta_3{}^2(t_c-t_2)+\mu\beta_2{}^2(t_2-t_1)+\mu\beta_1{}^2 t_1, \ t_c{\leq}t \ . \end{cases} \qquad \text{Eq.1.28}$$

Thus we see $\xi$ is continuous and piecewise linear, beginning at $\xi=0$ in $V_k{}^-$, and ending up as a constant in $V_k$ when $t{\geq}t_c$. A similar result applies in the case of the augmented parameter spaces $\mathbf{P_{1+2n,1}}$ which differs from $\mathbf{P_{1+2n}}$ by the addition of a $\mu>0$ parameter to incorporate the $\xi$ field. This is discussed at length in Ref.2, beginning with Eq.3.36.

Now you might be wondering why I require all of my model universes to end at some time $t_c$. After all, most inflationary universe models, such as Guth's original models (*see,* Ref. 10) go on forever without collapsing. (The cyclic universe models of Steinhardt & Turok, Ref.11 also go on forever.) Well, when it comes to cosmology, I take the action associated with my model universes seriously. Most people simply use the Lagrangian which generates the action as a source of the field equations, without ever considering the actual value assumed by the action. However, Feynman showed us in his path integral approach to Quantum Mechanics (Ref.14) that the action can be used to construct the wave function for various physical systems. In Ref.2, I present a method for using the action of a branch universe to determine how likely it is to be populated by matter in comparison with other branches. The closer the value of the action is to zero the more likely a branch is to be populated. While conversely the larger the value of a branch's action the less likely it is to be populated. Thus in the limiting case of an open universe, its action would (in general) be infinite, and hence it would have no chance of being populated. So this is why I have restricted my attention to universes of finite duration.

I just alluded to the problem of populating the universes of a multiverse.  So we shall now turn our attention to doing just that.

## Section 2: Cosmic Procreation–Populating the Multiverse

Presently we are working with the Lagrangian $L_T$ given in Eq.1.23 which is independent of matter terms.  Within the context of Lorentzian Cofinsler theory we were able to construct a simple multiverse, with each branch beginning at t=0 from a common 4-dimensional Lorentzian space $V_k^-$.  These branches passed through three de Sitter phases before returning to $V_k$ at some time $t \geq t_c$.  In the third section of Ref.2, I attempted to explain how each of these branches of the multiverse may become populated with matter.  After further excogitation on the subject I realize that my remarks in Ref.2, were, to put it politely, erroneous.  After all, my expertise lies in constructing field theories, not elementary particle theories.  Nevertheless, I thought that I would make another attempt to explain how the universes of a multiverse can become populated with particles.

To begin it would be naive to believe that if there really were a multiverse then the standard model of elementary particles would govern the behavior of the elementary particles in each branch of that multiverse.  We are all cognizant of the delicate balance between the masses and other parameters in the standard model that make our existence possible.  So these quantities should be permitted to assume other values in the various branches of the multiverse.

The equations of the standard model are derived from a Lagrangian.  So to begin with let us consider elementary particle theories that are derivable from a Lagrangian $\mathcal{L}$ which is such that if it needs a metric tensor in its formulation that metric comes from the FLRW space.  Assume that N parameters must be specified to determine a particle theory when using $\mathcal{L}$. We shall let $\mathbf{EP}\mathcal{L}(\mathbf{N}) \subset \mathbb{R}^N$ denote the domain of admissible parameters for the elementary particle theories which $\mathcal{L}$ can generate.    Thus each element of **EP𝔏(N)** determines an elementary particle theory, although it might not be the one suitable for a universe in which we could reside.   I shall now enlarge the parameter space for the branches of the multiverse from $P_{5,1}$ to $EP𝔏(N) \times P_{5,1}$.   Hence when we choose a parameter for a branch of the multiverse not only will we get a piecewise de Sitter universe consisting of three exponential branches beginning and ending in an open submanifold of $V_k$, but also an elementry particle theory to govern the behavior of the matter in that universe.

The next problem is: where does the matter that our elementary particle theories apply to come from?   My conjecture is that it comes from nothing in the following way.  I assume, for the present,  that when t<0 there are no elementary particle theories at work, and that the spatial part of $V_k^-$; *viz.,* $S^3$, is filled with primordial point particle that are identical except for the fact that some particles have positive energy E, and some have negative energy –E. (A precise value for E will not be required in this paper, but a logical choice would be the Planck energy.) I shall assume that these + and– energy particles reside as pairs at  points of $S^3$, and let *N* denote of the set of primordial particle pairs in $S^3$.   Since the total energy of each primordial pair is zero, the total energy of all the primordial particles residing on $S^3$ is zero.   Thus the total energy in the multiverse from particles when t<0 is zero.   I shall have more to say about where these primordial particles may have come from in the last section of this paper.

Now when t=0, I require that every one of the primordial particles must go individually (not in pairs) into some branch of the multiverse–how they choose is unknown and, as I mentioned above, is briefly discussed in Ref.2, using the value of the action integral over a branch.  It is highly unlikely that each branch will then be filled with "equal numbers" of + energy and – energy primordial particles, where, for the moment, let's assume that the total number of primordial particles that enter a branch is finite,  so we do not have to contend with equality of infinite sets.  I shall require that one law all of our elementary particle theories possesses when they become operative at t>0 is that +E and –E energy particles at the same point of $S^3$ annihilate each other, leaving nothing. Thus in a short period of time near t = $0^+$ all the + and – energy particles in a branch that can annihilate will do so leaving us with a preponderance of either + energy or – energy particles for the various particle theories to work with to create elementary particles for that branch. Hence each branch will be exclusively made from either + or – energy particles, but the total energy in the multiverse from elementary particles will be zero.

In Guth's original inflationary universe model (*see,* Guth Ref. 10) the universe does not expand immediately after t=0. There is a small period of time after t=0 which is required for thermal equilibrium of the particles to occur. I can arrange for that to happen in most of the branches of our multiverse by using the parameter space $\mathbf{P_{1,5,1}}$ defined by

$\mathbf{P_{1,5,1}} := \{(\varepsilon,(\beta_1,\alpha_2,\beta_2,\alpha_3,\beta_3),\mu) \in [0,\infty) \times (\mathbb{R}^+ \times \mathbb{R}^4) \times \mathbb{R}^+ | \exists\, t_c > t_2 > t_1 > \varepsilon \geq 0$ with $\exp(\beta_1(t_1-\varepsilon)) = \alpha_2\beta_2\exp(\beta_2 t_1)$,

$\alpha_2\beta_2\exp(\beta_2 t_2) = \alpha_3\beta_3\exp(\beta_3 t_2) > 1,\ \alpha_3\beta_3\exp(\beta_3 t_c) = 1$, where $\beta_i \neq \beta_{i+1}$, i=1,2\}$.     Eq.2.1

The $\varepsilon$ parameter in $\mathbf{P_{1,5,1}}$ corresponds to the delay after t=0, to allow thermal equilibrium to occur before expansion begins. The $\mu$ parameter is for the scalar field $\xi$. If $(\varepsilon,(\beta_1,\alpha_2,\beta_2,\alpha_3,\beta_3),\mu) \in \mathbf{P_{1,5,1}}$ then the scalar fields $\varphi$ and $\xi$ that it generates would be

$$\varphi := \begin{cases} t, & \text{if } t<\varepsilon, \\ \beta_1^{-1}\exp(\beta_1(t-\varepsilon)), & \text{if } \varepsilon \leq t < t_1 \\ \alpha_2\exp(\beta_2 t), & \text{if } t_1 \leq t < t_2 \\ \alpha_3\exp(\beta_3 t), & \text{if } t_2 \leq t < t_c, \\ t, & \text{if } t_c \leq t , \end{cases}$$     Eq.2.2

and

$$\xi := \begin{cases} 0, & \text{if } t<\varepsilon, \\ \mu\beta_1^2(t-\varepsilon), & \varepsilon \leq t < t_1 \\ \mu\beta_2^2(t-t_1)+\mu\beta_1^2(t_1-\varepsilon), & t_1 \leq t < t_2 \\ \mu\beta_3^2(t-t_2)+\mu\beta_2^2(t_2-t_1)+\mu\beta_1^2(t_1-\varepsilon), & t_2 \leq t < t_c \\ \mu\beta_3^2(t_c-t_2)+\mu\beta_2^2(t_2-t_1)+\mu\beta_1^2(t_1-\varepsilon), & t_c \leq t . \end{cases}$$     Eq.2.3

From Eqs.2.1and 2.2 we see that the universe determined by $\varphi$ does not begin to expand exponentially at t=0 (if ε >0), but remains a $V_k$ until t=ε. So the multiverse $\mathbf{MV_k(EP\mathcal{L}(N) \times P_{1,5,1})}$ provides us with a multiverse in which each branch is populated by EPs (:= elementary particles) governed by an N parameter particle theory generated by the matter Lagrangian $\mathcal{L}$.

One technical problem here is the cardinality of $\mathit{N}$, denoted by card($\mathit{N}$), where recall that $\mathit{N}$ is the set of of primordial particle pairs in $S^3$ before time multifurcates at t=0. Rather than get bogged down with the problem of how many angels fit on the head of a pin, I shall assume that card($\mathit{N}$) is so large that no matter how big the multiverse is that we construct, there are at least "a lot" (and I would not mind if there were an infinite number) of EPs in each branch of the multiverse. *E.g.,* let us suppose that card($\mathit{N}$) = $\mathit{c}$ (the cardinality of the continuum, which is card($\mathbb{R}$)). Using the fact that if $\mathit{a}$ is an infinite cardinal number then $\mathit{a \cdot a} = \mathit{a}$ (*see*, section 24 in R. Halmos, Ref. 15 for this result and other terminology on cardinal numbers and their properties, such as: if A and B are sets then card(A)·card(B) :=card(A×B)), it is relatively easy to show that the cardinality of the set of branches of $\mathbf{MV_k(EP\mathcal{L}(N) \times P_{1,5,1})}$ is $\mathit{c}$. Consequently when card($\mathit{N}$) = $\mathit{c}$ each branch of $\mathbf{MV_k(EP\mathcal{L}(N)xP_{1,5,1})}$ could have $\mathit{c}$ primordial particles of energy +E or −E after t=0. Thus it is not unreasonable to assume that whatever card($\mathit{N}$) might be, there is some value for this quantity that will guarantee that when these primordial particles enter a branch at t=0, after a great deal on annihilation, we shall be left with "a lot"of either all +E or all −E primordial particles. I shall discuss this issue again later in this paper.

Thus far we have been dealing with the parameter spaces $\mathbf{P_5}$, $\mathbf{P_{5,1}}$ and $\mathbf{P_{1,5,1}}$ when building universes and multiverses. These parameter spaces led to universes with three exponential phases. As I mentioned previously, in Ref.2, I introduced the parameter spaces $\mathbf{P_{1+2n}}$ (n≥2) which provide n+1 exponential phases. $\mathbf{P_{1+2n,1}}$ and $\mathbf{P_{1,1+2n,1}}$ could be built from $\mathbf{P_{1+2n}}$ in the obvious way using Eqs.1.20 and 2.1 as models. Consequently we could build a multiverse $\mathbf{MV_k(EP\mathcal{L}(N) \times P_{1,1+2,n,1})}$ ∀

n≥2. All of these multiverses would have the same tail; *viz.,* $V_k^-$. So if we take the formal union of all these spaces for n≥2, we could then glue their tails together to form the four dimensional megamultiverse starting from $V_k^-$ with particle Lagrangian $\mathscr{L}$ defined by

$$\mathbf{MMV_k(EP\mathscr{L}(N))} := \bigcup\nolimits_{n \geq 2} \mathbf{MV_k(EP\mathscr{L}(N)) \times P_{1,1+2n,1}} \, / \sim \qquad \text{Eq.2.4}$$

where ~ is the equivalence relation that glues the tails together. A global time function t, scalar field φ, and Cofinsler function f on $\mathbf{MMV_k(EP\mathscr{L}(N))}$ and $\mathbf{T^*MMV_k(EP\mathscr{L}(N))}$ could be defined in the obvious way to provide us with a Lorentzian metric tensor g which was everywhere continuous and piecewise of class $C^\infty$. But we are not yet done in our construction of the ultimate multiverse.

Let **Lag** denote the set of all possible Lagrangians from which elementary particle theories can be made. $\forall \, \mathscr{L} \in \textbf{\textit{Lag}}$ the megamultiverse defined in Eq.2.4 has the same tail. Thus we can glue these tails together to obtain the ultimate four dimensional megamultiverse starting from $V_k^-$

$$\mathbf{UMMV_k} := \bigcup\nolimits_{\mathscr{L} \in \textit{Lag}} \mathbf{MMV_k(EP(\mathscr{L}(N))} \, / \sim' \qquad \text{Eq.2.5}$$

where ~' is the equivalence relation which identifies points on the tails of $\mathbf{MMV_k(EP(\mathscr{L}(N))}$. The four-dimensional manifold $\mathbf{UMMV_k}$ has a global time function t that multifurcates at t=0, a scalar field φ and a Lorentzian Cofinsler function f for $\mathbf{T^*UMMV_k}$. As usual the Lorentzian metric tensor defined by φ and f is continuous everywhere and piecewise of class $C^\infty$.

This is all well and good and gives us great insight into how enormous the multiverse could be if it had the form I described. But it still suffers from Nietzsche's complaint about action principles, which essentially was "the principle of least action and greatest stupidity." The stupidity consists in our not knowing why Nature would choose an action principle to select equations to describe the world, or why there should even be equations which describe the behavior of matter. Well, for us, it is obvious why Nature needs laws to govern elementary particles, since if there were no such fixed laws in at least one universe, we would not be here. But the multiverses that I have presented allow the possibility of lawless, and partially lawless branches. To see why note that I did not stipulate that the Lagrangians in **_Lag_** had to be scalar densities. So some could have explicit time dependence. *E.g.,* think of the usual Lagrangian of the standard model, $\mathscr{L}_{SM}$, and let h=h(t) be a smooth real valued function defined on $\mathbb{R}$, that varies between 0 and 1. Assume that $0_h := \{ t | h(t)=0, t>0 \}$ and $1_h := \{ t | h(t)=1, t>0 \}$ both have nonempty interiors. If we now consider the Lagrangian $h \cdot \mathscr{L}_{SM}$ it will provide us with a megamultiverse $\mathbf{MMV_k(h \cdot \mathscr{L}_{SM}(N))}$ in which each branch will contain regions where the standard model of elementary particles is operating (with h(t)=1), and other regions of complete chaos (with h(t)=0). So in a certain sense $\mathbf{UMMV_k}$ addresses Nietzsche's complaint since it is possible to use a variational principle to construct total lawless chaos. We just do not live there now, although it could happen in our future, in which case we would vanish.

Now many religious people may think that this idea of the $\mathbf{UMMV_k}$ is just way to complex. They feel that God wants things to be simple. That is probably what Maupertuis thought when he introduced the principle of least action in the 1740s. But I contend that $\mathbf{UMMV_k}$ is the simple way. Instead of God tinkering around with Lagrangians and parameters, God took the path of least resistance and simply said: "Let There Be Everything. And I'll come back in a few days (perhaps six) and see what I got."

The megamultiverses $\mathbf{MMV_k(\mathscr{L})}$ and ultramegamultiverse $\mathbf{UMMV_k}$ are analogous to what Tegmark, Ref.16, calls multiverses of type I and type II. However, each of Tegmark's multiverses last forever, while mine are such that each branch has only a finite lifetime (even though each branch is perfectly capable of bouncing and starting its cycle over again forever). In addition the multiverses I have introduced do not require some form of matter at the outset to create the individual universes. The universes are created by $\varphi$, which when combined with $\xi$, satisfy source-free field equations. Moreover, the total energy required for all the particles in the multiverse is zero.

**Section 3: What Happened to Gravity?**

That is a good question.  The elementary particle theories we considered in the previous section involved a Lagrangian that did not incorporate a term involving gravity.  Any time a metric tensor was required in those Lagrangians the metric of the enveloping FLRW space involving the scalar field φ was used.  But nowhere do I attempt to explain why we are held firmly to the earth. To do that I believe that what we require is a quantum theory of gravity.  So what I shall do is discuss the rudiments of how such a theory might be incorporated within the framework of the multiverses I have constructed.

When t<0 all of our universes are spatially an $S_k$ for some fixed value of k>0.  Let us now choose one value of k to work with.  My choice is the value of k which gives $1L^{*3}$ as the volume of the 3-spaces, $S_k$, when t<0, where $L^*$ is the reduced Planck length, $L^* := \hbar^{\frac{1}{2}}$, when c=G=1.  If $R_0$ is the radius of this 3-sphere, then $k = R_0^{-2}$, and the volume of this sphere is $2\pi^2 R_0^3$.  So if we choose $k = (4\pi^4)^{1/3}$ then when t<0 the volume of the t = constant slices will be $1L^{*3}$.  I shall call this $S_k$, $S_i$, for the initial 3-space of the multiverse when t<0.  Throughout the remainder of this paper I shall confine our attention to $S_i$, although much of what I say shall have an obvious extension to $S_k$.

In keeping with this idea of choosing one value of k to work with, let's also assume that for each Lagrangian $\mathcal{L} \in \textbf{\textit{Lag}}$, c, G and $\hbar$, have the same value.  Hence all branches of the ultramegamultiverse will have the same values for the Planck time, $T^*$, and Planck length, $L^*$.

From Einstein's work we suspect that it should be possible to study gravity as a feature of the geometry of spacetime.  This suggests that a quantum theory of gravity should be constructed from a quantum theory of geometry.  So our quantum geometry begins quite simply with a single cell of space which is $S_i$.  As space begins to expand exponentially in a branch universe at some time greater than 0 (dependent upon the branch one is in) the original $S_i$ will double in volume.  At that time I shall require that the original cell divides into 2 connected cells of volume $1L*^3$, with a common 2-dimensional boundary. This cell division process will continue as long as space is expanding, thereby populating each branch of the multiverse with cells which have the same volume at each instant of time, and that volume will range from 1 to 2 $L*^3$. Now when space begins to contract a cell with volume $L*^3$ will eventually shrink down to a volume of ½ $L*^3$. At that time two cells that have a common 2-dimensional boundary will merge to form one cell of volume $L*^3$. This merging of adjoining cells will continue as the universe branch shrinks until only one cell of volume $L*^3$ remains, our original $S_i$. In these models the basic quantum cells of space have a volume ranging from ½$L*^3$ to 2$L*^3$. Exactly what these cells might look like is immaterial since they can not be seen.

All of the cells after the first split and before the last merger will be homeomorphic to the 3-cube, but when they start dividing the borders between touching cells can be either points, 1-dimensional manifolds or 2-dimensional manifolds with boundaries, and the situation will be chaotic. I shall assume that all of the cells have clocks synchronized with the global time function of the multiverse, and these clocks measure time in discrete integer units of the reduced Planck time $T* = \hbar^{½}$, when c=G=1. Note that depending upon the universe branch it is possible for the cells of space to be dividing (or combining) in less time then the discrete ticks of the universe clock.

Now for EPs (:=elementary particles) in the multiverse. I shall assume EPs are points which ride on the surfaces of the cells, and distribute themselves "uniformly" over the surface. These point particles will be either massless or not. The requirement is that at each tick of the universal clock any massless EP in a cell C must move to some adjoining cell C' which makes contact with C in such a way that $\partial C \cap \partial C'$ has a two-dimensional interior at that moment. I call two cells C and C' that meet in the manner just described as contiguous. Hence contiguous cells can not just meet at one point (a vertex) or along an edge.

At this time it should be noted that since the number of cells in a universe increases as it expands, we should expect that the entropy of a branch is non-decreasing while it is expanding.  On the other hand, when a universe branch is contracting, the number of cells in it is decreasing, and hence I expect that when a universe branch is in a contracting phase the entropy will be non-increasing.  This contradicts the usual second law of thermodynamics which was developed in a virtually flat expanding universe, and hence need not apply in more general situations.  In addition we see that this second "law" of thermodynamics changes from branch to branch of the multiverse, with the direction of the inequality depending upon whether $\varphi''$ is positive or negative.

If we hope to do geometry I need to introduce a notion of distance.  One way this can be done is as follows.  To specify the distance between two cells $C_1$ and $C_2$ in the same branch of a multiverse we need to specify a time, since the number of cells in the universe is changing in time, as are $C_1$ and $C_2$.  What we could do is take a "snapshot" of the universe at a given moment and then look at the various paths one could take through contiguous cells to get from $C_1$ to $C_2$.  These paths would be like strings of beads, with the contiguous cells being the beads.  I define the distance from $C_1$ to $C_2$ at time $t_0$, denoted $d(t_0;C_1,C_2)$ to be L* times the number of cells in that path from $C_1$ to $C_2$ which has the fewest number of cells in it.

Things get a bit more difficult when we try to define the length of a path $\gamma = \gamma(t)$, $0 \leq a \leq t \leq b$, which could be the trajectory of an EP through U, a branch of the multiverse.  We shall assume that $\forall\, t \in [a,b]$, $\gamma(t)$ denotes a cell, and that $\gamma$ only passes through contiguous cells.  Due to the fact that space may be subdividing (or combining) many times in a time < T*, much of the "movement" of $\gamma(t)$ may simply be due to changes in the enveloping space.  Let $\#\gamma([a,b])$ denote the number of contiguous cells that $\gamma$ passes through as t goes from a to b. In this definition if, say, $\gamma((c,d))$ lies in some cell, $a \leq c < d \leq b$, and then at t=d the cell either splits or combines, that will add a 1 to $\#\gamma$, because the path has entered a new cell and left the old one.   I then define the length of γ to be L*·#γ([a,b]).

I would now like to introduce a notion of distance between the pair $(t_1,C_1)$, consisting of a cell $C_1$ at time $t_1$, and a second pair $(t_2,C_2)$, where $t_1 < t_2$, and $C_1$, $C_2 \in U$ (we treated the $t_1 = t_2$ case above).   To that end let $\gamma = \gamma(t)$, $t_1 \leq t \leq t_2$, be a path in U, the universe branch the cells are in.   As above, $\forall\, t \in [t_1,t_2]$ γ(t) is a cell in U at time t, and nothing more–it is not a point in U.   I define the distance from $(t_1,C_1)$ to $(t_2,C_2)$ by

$$d((t_1,C_1),(t_2,C_2)) := L^* \cdot min\{\#\gamma([t_1,t_2]) | \gamma:[t_1,t_2] \to U \text{ is a path from } (t_1,C_1) \text{ to } (t_2,C_2)\}. \quad \text{Eq.3.1}$$

We can use these notions of distance to define the "proper distance" along a curve $\gamma = \gamma(t)$, $t \in [t_1,t_2]$, or between two cells $C_1$ and $C_2$ at times $t_1$ and $t_2$, denoted by $s(\gamma,[t_1,t_2])$ and $s((t_1,C_1),(t,C_2))$ by

$$s^2(\gamma,[t_1,t_2]) := -(t_2-t_1)^2 + \{L^* \cdot \#\gamma([t_1,t_2])\}^2$$

and

$$s^2((t_1,C_1),(t_2,C_2)) := -(t_2-t_1)^2 + d^2((t_1,C_1),(t_2,C_2))\ .$$

You should note that the notions of distance just introduced are intimately connected with the FLRW geometry of U since it is that geometry which is being employed to determine the volume of the cells.   In addition if γ were the trajectory of a massless EP, then $s(\gamma,[t_1,t_2])$ will in general be greater than 0, due to the expansion or contraction of the ambient FLRW space.

Gravity enters these branch universes through Gs (:= gravitons) which are responsible for gravitational effects.   But what should these effects be?  The obvious effect is that Gs cause EPs to move toward one another.   However, that effect would not change the geometry of the branch universes.   So we require Gs to cause attraction and also leave their imprint on the geometry.   I suggest that one way of accomplishing this is to stipulate that a G is an "order" emanated by an EP to the cells of space which travels like a massless particle.   The rate at which an EP emits Gs will not be required here but should be directly proportional to the energy of the EP at the time it emits  Gs.

When a G arrives at a cell containing matter it proceeds no further and it directs that cell to immediately move all of the matter in it at the time the G arrived into the cell that the G just left. After doing that the cell is required to shut down for 1 tick of the clock. While a cell is "off line" neither it, nor any of the cells it splits into or combines with, will be permitted to allow anything to enter or leave until after 1 tick has passed. The requirement that a cell and its immediate descendants or combinations must shut down for one T* affects the geometry of the universe. To see this consider the pairs $(t_1, C_1)$ and $(t_2, C_2)$, and let $C_0$ be a cell that shuts down at some time $t_0$, $t_1 < t_0 < t_2$. Suppose that the path $\gamma = \gamma(t)$, $t_1 \leq t \leq t_2$, would be the shortest path from $(t_1, C_1)$ to $(t_2, C_2)$ if $C_0$ did not shut down at time $t_0$; *i.e.,* if gravity were not present. But since gravity is present the shortest path (assuming there were not several shortest paths from $(t_1, C_1)$ to $(t_2, C_2)$) must now circumvent $(t_0, C_0)$ and hence be longer. Thus gravitational effects, as just defined, change the geometry of the universe by in general making travel distances, and hence travel times between cells longer.

For clarification I would like to state that when I refer to matter, I mean any EP, whether it is massive or massless, including all of the force particles except gravitons. Gravitons owe their existence to what I am calling matter, and provide a link between matter and geometry.

In view of my present definition of Gs and their behavior it seems difficult to believe that matter in the early universe could even move since many cells would be off line for much of the time. To overcome this absurd situation I shall assume that a G has to move through at least one empty cell before it becomes "activated." And if it encounters an EP in an adjacent cell it is vitiated; *i.e.,* it disappears and is no longer capable of providing directions to cells. This restriction will permit the matter in the universe to expand into U and it also has implications for the gravitational collapse of any celestial body. This is so since it helps prevent "singularities" from forming in which gravity causes all of the mass of a collapsing body into one cell of space. To see why this is so suppose we have a connected region R = R(t), which is a function of time, with R(t) having more than one cell. Assume that at time $t_1$ all the cells in $R(t_1)$ are occupied by at least one EP. I shall say that a pair $(t_1,C_1)$ is in the interior of $R(t_1)$, denoted int($R(t_1)$), if every contiguous cell of $(t_1,C_1)$ is also in $R(t_1)$. Gravity does not exist in int($R(t_1)$) since there can be no active gravitons in that region. Hence the int($R(t_1)$) can not collapse down to one cell because of its own "weight." We can think of int($R(t_1)$) as a gravitationally superconducting medium since gravity offers no restriction to the movement of EPs in that region. This behavior is somewhat similar to the asymptotic freeness of the strong force.

The boundary of $R(t_1)$, denoted by $\partial R(t_1)$, consists of those cells of $R(t_1)$ which have at least one cell which is not in int($R(t_1)$). Due to the above remarks we know that the gravitational effects of $R(t_1)$ are determined by the cells of $\partial R(t_1)$. In a gravitational collapse of a celestial object into a region R(t) once things have settled down at time $t=t_1$, all the cells on $\partial R(t_1)$ will have neighboring cells that are vacant. Thus the gravitational effects of the collapsed object will be determined entirely by the EPs, contained on $\partial R(t_1)$. This observation would provide an experimental test of these ideas. The test would require knowing what the mass was of the collapsing object after it finished ejecting mass and began its inextricable collapse through its event horizon. Let us call this mass of the collapsing object, $m_c$, which would then be the inertial mass of the black hole. Then a short while after the object disappears through the event horizon and things settle down, we can determine from the external gravitational field what the mass of the object causing that external field is. Call that mass, $m_b$, for the gravitational mass of the black hole. The theory that I have presented above predicts that $m_b < m_c$, and that the event horizon should have shrunk accordingly. So for black holes this theory implies that the gravitational mass is less than the inertial mass.

In the previous section I introduced the scalar field $\xi$. The primary reason for its existence at that time was to dampen exponential oscillations of my model universes, by causing the action to increase every time the universe experiences a "bump," or a "dip" during the course of its evolution. Hence to minimize the action you need to minimize the oscillations, and that is essentially what $\xi$ helps to achieve. I want to think of $\xi$ as creating what we currently regard as the effects of dark matter. To do that I need to explain how $\xi$'s quanta interact gravitational with EPs in the present context. I suggest that oneway of doing that goes as follows.

From Eq.2.3 we see that $\xi = \xi(t)$ permeates all space and it is piecewise linear in t. Let's assume that $\xi$ produces gravitons, and let $G_\xi$ be one of $\xi$'s gravitons. $G_\xi$ behaves differently then the other gravitons produced by EPs. I shall assume that if $G_\xi$ enters a cell that only has EPs in it, then it will due nothing and continue onward behaving like a massless particle. However, if a $G_\xi$ is in a cell with matter when a regular graviton G enters, then it is activated and adds its strength to G, thereby enabling G to pull EPs into the cell it just left with a bit more strength. This helps to explain why you only see dark matter effects near regions occupied by matter, because the $\xi$ field's gravitons need both matter and matter gravitons to come to life.

It should be noted that if I had allowed the gravitons of the $\xi$ field to behave like the matter gravitons, then the $\xi$ field would have tried to pull galaxies apart. This is so since there is much more $\xi$ field located outside of galaxies then within them, and thus matter within galaxies would feel a force trying to pull it toward empty space outside the galaxies.

Although the gravitational theory I have briefly outlined in this section is far from complete, one thing is certain and that is that "local gravity" has no affect upon the expansion or contraction of the universes that comprise the multiverse. That behavior is completely determined by $\varphi$ and $\xi$. Gravity along with the other fields relating to the forces between elementary particles generated by the particle Lagrangians simply carry out their interactions in the arena provided by the ambient FLRW metric generated by $\varphi$ through Lorentzian Cofinsler geometry.

During the early phase of a universe, when matter is more concentrated, the gravitational effects can give the illusion that the universe is not expanding that rapidly. It is only after matter starts to form larger bodies, and these bodies get distributed over larger distances by the ambient expansion of the universe, that the local gravitational effects diminish, and are eventually overwhelmed by the expansion of the universe generated by the φ field. Until this occurs it may appear as though the expansion of the universe is not exponential, even though there really has been no overt change in the exponential expansion of the ambient universe. This results from our inability to actually see space itself. We can only observe matter, and radiation (which is part of matter according to my terminology), in space, and from those observations we try to infer what space might be doing.

The best experimental evidence we have that gravity is not responsible for the ambient geometry of a universe is that currently our universe's matter distribution does not appear to be homogeneous and isotropic on any scale. While assuming that the large scale geometry of the universe is of the FLRW type does seem to explain the expansion, and possibly future contraction, of our universe. By way of analogy, if the universe were a bus, then gravity and matter are not driving that bus. They are just sitting in the back, looking out the window, enjoying the ride, not knowing where they are going or when they are going to get there. φ and ξ (as coupled in the Lagrangian $L_T$) are driving the bus. If anything we should regard the ambient FLRW metric determined by φ and ξ as providing asymptotic boundary conditions for the local metric tensor $g_{ab}$ that gravity generates. A similar remark pertains to φ and ξ as providing asymptotic boundary conditions for any scalar-tensor or scalar-scalar-tensor field theory that tries to describe gravity locally. Perhaps such theories should be regarded as providing yet another layer of parameters to enlarge the multiverse, just as the different Lagrangians in **Lag** did. It seems like the multiverse's possible complexity is like a snowball rolling downhill, getting ever larger as it goes.

**Section 4: Concluding Remarks**

Earlier I mentioned that in Ref.2 I investigated the k=0, FLRW spaces in the context of Lorentzian Cofinsler spaces and scalar-scalar field theories. While doing so I realized that one problem with those theories is how matter enters the branch universes at t=0. Each t=constant slice for t<0 in these models is $\mathbb{R}^3$. If $\mathbb{R}^3$ had $N$ primordial particle pairs, then as t passes through 0 these primordial particles would randomly populate the $\mathbb{R}^3$s of the various multiverse branches. So that even if we have some delay before each universe branch begins to expand it might be very difficult to get all the E and ‑E primordial particles to annihilate, and it would be even harder to achieve thermal equilibrium. That is the main reason I chose to consider the k>0, FLRW spaces here. But while working on this paper I realized that there exists a very easy way to get around the k=0 particle problem. All we need do is to assume that when t<0 all of the primordial particle pairs are not randomly attached to points of $\mathbb{R}^3$, but are all in a cube of edge length L* centered at (0,0,0). As t passes through 0 all of the primordial particles do their usual scattering among the branches of the k=0 multiverse, but they are required to remain in the cell centered at (0,0,0). They remain in that cell until the universe begins to expand after the branch's initial delay to allows EPs to form and achieve thermal equilibrium. For these models when t=0 the initial $\mathbb{R}^3$ would be built from cubical cells of edge length L*. When the universe branches begin expanding, the edge length of a cell will eventually become 2L*. At that time I shall require the cell to split into 8 contiguous cubical cells of edge length L* that share common 2-dimensional square walls where they make contact. This continues as long as the universe branch expands. When the branch begins to contract, the edge length of a cell will eventually shrink to ½L*. At that time I shall require eight contiguous cells to merge into one cubical cell of edge length L*. It is easier to envision what is going on with the basic cell structure in these k=0 models then it is for the k>0 models, where the cells break up into ever weirder shapes as time evolves.

However, there is one problem that the k=0 models have that the k>0 models do not have. For the k>0 models, when the universe collapses back to the primordial $S_k$ at t=$t_c$ all the matter in the branch is now in that $S_k$. It is not at all clear if something like this could happen in the k=0 models. For example, when some of the models based on $\mathbf{P_{1,5,1}}$ (*see*, Eq.2.1) enter the exponential $\beta_2$ coast stage with $\beta_2 \approx 0^+$ the cells of space are hardly expanding at all, but various EPs could be shooting off into empty space away from (0,0,0) at speeds close or equal to light. It is very hard to believe that in general when the universe branch begins to contract all of these wandering EPs will make it back to the original cell at t=$t_c$, even when the effects of the local gravitational field are taken into account. Nevertheless I feel that the k=0 case warrants further investigation since it is much easier to work with $\mathbb{R}^3$ and its Euclidean geometry then it is to work with $S^3$. This is especially so when we try to explain how gravitons change the momentum of EPs they interact with, which I vainly try to do with non-inflationary models in Ref. 17.

In the context of all the various multiverses I have discussed in this paper we shall find the basic structure of our universe, the φ, ξ and matter Lagrangian $\mathcal{L}_{SM}$, in some branch. However, to get our exact universe we really need more than φ, ξ and $\mathcal{L}_{SM}$. We need all the initial energy from primordial particles to make just the right mix of EPs and have them interact over time in just the right way to make a planet earth in which say, an asteroid hits at just the right time to make gasoline for our cars today. That seems extremely far fetched, and yet here we are. The problem here is that the model with precisely our φ, ξ, and $\mathcal{L}_{SM}$ only occurs once in all of my multiverse and ultramegamultiverse models. Now earlier I briefly mentioned that each of the branch universes I constructed did not have to end in a $V_i$ (or $V_k$), at time t=$t_c$, but could "bounce" and just start all over again using essentially the same functions $\varphi$, $\xi$ and $\mathcal{L}_{SM}$ time translated to the next epoch. Logically this seems to make the most sense and leads to an interesting mathematical problem. If, say, there were only a finite number of primordial particles with energy E in a branch shortly after its outset, then could every possible particle history of all the EPs constructed from these primordial particles occur, somewhere in the countably infinite number of universes built within their lifetime of length $t_c$? I am not sure, but I would not be surprised if it did. However, my objective in constructing the **UMMV$_i$** was not to find us somewhere in it. The purpose was to try to explain inanimate reality, and perhaps somewhere in that enormous reality, living, sentient beings like us will occur, with some sort of bizarre history explaining how they came into existence in the universe branch they are in. We are not icing on the **UMMV$_i$** cake, but some infinitesimal spec crawling around in it, trying to decipher where we are, and how we might have got here. Some cosmological multiverse models (*c.f.,* Tegmark, Ref 16) claim that there are going to be an infinite number of copies of us scattered through the multiverse. I doubt if that will happen in my models. It will most likely be one and done, for exactly us. However, there will be an infinite number of universes with $\varphi$ and $\xi$ close to ours using $\mathcal{L}_{SM}$ as their matter Lagrangian, and they most likely will have sentient life somewhere in their histories.

A major complaint about multiverse theories is that since all of the various branches do not directly interact with one another, there is no way to tell if they are there. So does the multiverse really provide us with any information about our universe that we could not obtain in some other manner? Well, one insight that my theory provides is an explanation of where all the matter in our universe came from to begin with, *viz.,* nothing. And this would not be possible without the multiverse's existence. The theory does not also require the distribution of matter to be homogeneous and isotropic on any scale, which presently seems to be the case in our universe.

As I have constructed my multiverses there is no direct interaction between the various branch universes when t>0. However, in Ref.2, I demonstrate that it is possible to build a time manifold from an arbitrary parameter space $P_n$ which is such that for t>0 we can travel in time from one time coordinate domain $U_p$ to another $U_{p'}$, when p, p'$\in P_n$ are distinct. Thus it is possible to build multiverses in which time travel between branches is mathematically possible. But in the construction developed in Ref 2 (*see,* my remarks in the paragraph following the one that contains Eq.3.6 , in that paper) the time travel is not local. What happens is that time bifurcates at some time t=$\tau$>0 and space splits into two segments similar to the way space splits at t=0 into multiple universes. One space carries on as $U_p$ and the other moves off and merges with $U_{p'}$. It might be possible to construct more local time travel, say starting within a black hole where you are dealing with a portion of space which has somewhat isolated itself from the rest of the branch, but I have not been able to accomplish that. In any case you would need some amazing space suit that protects you as you go from one FLRW geometry to another.

Another question people have about most multiverses and the big bang (the Steinhardt, Turok cyclic model of Ref.11 excepted), is what was going on before t=0? *E.g.,* in my **UMMV**$_i$ there is an infinite amount of time before anything happens at t=0. My explanation for what happened during this period is quite simple. Recall that to construct matter in the branches of **MMV**$_i$, I required there to be *N* primordial particle pairs, where card(*N*) is an infinite cardinal number. Where did those primordial particle pairs come from? Well, recall that in standard QFT in Minkowski space, particle-antiparticle pairs are always being created somewhere in the vacuum. This occurs when the pair borrows an amount of energy 2$\mathscr{E}$ from the vacuum, and that enables them to come into existence for a time $\tau$, which is such that 2$\mathscr{E}\tau<\hbar$. Now I previously stipulated that there was no quantum field theory at work governing the primordial particles when t<0. I shall now introduce one, and it is a very simple one. It does only one thing and that is to create primordial particle pairs of energy E and −E at points of $S_i$ when t<0, which is where it acts. Since the energy borrowed from the primordial vacuum is zero, these pairs can live forever since they trivially satisfy the condition that $2\mathscr{E}\tau<\hbar$. Now let us suppose that when t<0, time is quantized in units of T*, and that the PQFT (:=primordial QFT) can only produce particle pairs during quantums of time, and not continuously. The cardinality of the set of T* units of time <0, is $\aleph_0$. Hence if the PQFT only produced a finite number of particle pairs during each quantum of time, the cardinality $N$ of the set of produced pairs would be $\aleph_0$, which is the cardinality of the set of natural numbers, $\mathbf{Z}^+ := \{1,2,3,...\}$. Thus when t<0, the **MMV$_i$** would be preoccupied with building the particles necessary to fill the branches of the various universes which come into existence when t>0. And it would need an infinite amount of time to do just that. So we have just answered a version of the question: which came first, the chicken or the egg? The answer here is the egg. The egg being the 3-space $S_i$ of volume L*$^3$, when t<0. This egg gestated over an infinite amount of time, and when it finally contained $\aleph_0$ baby chickens, *i.e.,* primordial particle pairs at t=0, it gave birth to the particles that reside in the various branches of **MMV$_i$** when t>0.

Of course, during this gestation period, $\aleph_0$ primordial particle pairs were produced when t=−1T*, or t=−$10^6$T*, or t=−$(10^{100!})$!T*, or t= any negative integer times T*. So any time after $\aleph_0$ primordial particle pairs finally appeared in the egg $S_i$ could be taken as t=0 for **MMV$_i$**, with the choice of t=0 previously made for all our multiverses being good enough.

It is interesting to note that even though matter plays no roll in the construction of the geometry of any of the multiveses I have constructed, it plays a crucial roll in their existence. The multiverses essentially have to wait until the primordial particle pairs reach the "critical mass" with cardinality $\aleph_0$ at t=0, so that time can then multifurcate, permitting all of the branch universes to come into existence. Thus matter is really the main attraction in the multiverse even though it physically occupies only a small portion of it. So although matter and its accompanying gravitational field are not driving the bus, the bus can not start until they get aboard.

Now if we wanted $N$ to be $c$ (or larger) with time still quantized in units to T*, then we would need the PQFT to produce that number of particles at some instant of time, and hence we would not require an infinite amount of time to produce a $c$'s worth of particles. Thus we would still be left with the problem of trying to figure out what the **MMV**$_i$ was doing during all of the rest of the time <0, since it did not need it to make primordial particle pairs. Hence I shall stick with the first option in which the PQFT in the **MMV**$_i$ produces $\aleph_0$ primordial particles during the infinite amount of time <0. But this option begs the question which is: if $N = \aleph_0$, then will there be enough particles available for the **UMMV**$_i$ that I have constructed? Clearly not unless some restrictions are imposed on *Lag* and the range of the parameters in $\mathbf{P}_{1,1+2n,1}$. Fortunately $\mathbf{P}_{1,1+2n,1}$ can be "rationalized" without much loss of generality in the following manner.

Let $\mathbf{Q}$ denote the set of rational numbers, with $\mathbf{Q}^+$ denoting the positive rational numbers and $\mathbf{Q}^{\geq 0}$ the non-negative rational numbers. I now define the "rationalized" version of $\mathbf{P}_{1,5,1}$, denoted $\langle \mathbf{P}_{1,5,1} \rangle$ by

$$\langle \mathbf{P}_{1,5,1} \rangle := \{ \ (\varepsilon, \ (\beta_1,\alpha_2, \ \beta_2,\alpha_3,\beta_3), \ \mu) \in \mathbf{Q}^{\geq 0} \times (\mathbf{Q}^+ \times \mathbf{Q}^4) \times \mathbf{Q}^+ \mid \exists \text{ real numbers } t_c > t_2 > t_1 > 0 \text{ with}$$
$$\exp(\beta_1(t_1 - \varepsilon)) = \alpha_2\beta_2\exp(\beta_2 t_1), \ \alpha_2\beta_2\exp(\beta_2 t_2) = \alpha_3\beta_3\exp(\beta_3 t_2) > 1 \text{ and } \alpha_3\beta_3\exp(\beta_3 t_c) = 1,$$
$$\text{where} \quad \beta_i \neq \beta_{i+1}, \ i = 1,2 \}. \qquad\qquad \text{Eq.4.1}$$

Since card $\mathbf{Q} = \aleph_0$, and $\forall \ n \in \mathbf{Z}^+$ card $\mathbf{Q}^n = \aleph_0$, we have card $\langle \mathbf{P}_{1,5,1} \rangle = \aleph_0$. The scalar fields $\varphi$ and $\xi$ generated by $\langle \mathbf{P}_{1,5,1} \rangle$ would be defined just as they were in Eqs.2.2 and 2.3. We would also compute the value of the action associated with $L_T$ in the usual manner.

If $\mathscr{L}$ is a matter Lagrangian which requires N parameters to specify a field theory, then we previously let $\mathbf{EP\mathscr{L}(N)}$ denote the subset of $\mathbb{R}^N$ which is the domain of these parameters. We would replace $\mathbf{EP\mathscr{L}(N)}$ with only rational values for the parameters, denoted by $\langle\mathbf{EP\mathscr{L}(N)}\rangle{\subset}\mathbf{Q}^N$, which has cardinality $\leq\aleph_0$. So for each choice of an element of the set $\langle\mathbf{EP\mathscr{L}(N)}\rangle{\times}\langle\mathbf{P_{1,5,1}}\rangle$, which has cardinality $\aleph_0$, we can construct one of our piecewise de Sitter universes ending in a $V_i$, along with a particle theory to govern matter in that universe. This would provide us with the "rationalized" multiverse $\langle\mathbf{MV_i(EP\mathscr{L}(N){\times}P_{1,5,1})}\rangle$.

We can build $\langle\mathbf{P_{1,1+2n,1}}\rangle$ in a manner to similar to $\langle\mathbf{P_{1,5,1}}\rangle$ given in Eq.4.1. This space would also have cardinality $\aleph_0$. The "rationalized" megamultiverse determined by the Lagrangian $\mathscr{L}$ is defined in a manner analogous to Eq.2.4 as

$$\langle\mathbf{MMV_i(EP\mathscr{L}(N))}\rangle := \bigcup_{n\geq2} \langle\mathbf{MV_i(EP\mathscr{L}(N){\times}P_{1,1+2n,1})}\rangle / \sim ,$$

where $\sim$ is the equivalence relation which glues the tail of multiverses together. Since a countably infinite union of countably infinite sets is countably infinite, we have the cardinality of the number of branches in $\langle\mathbf{MMV_i(EP\mathscr{L}(N))}\rangle$ equal to $\aleph_0$.

Thus far we have been doing pretty good with our "rationalization" program. Where we shall hit a major road bump is when we take a union of the spaces $\langle\mathbf{MMV_i(EP\mathscr{L}(N))}\rangle$ as $\mathscr{L}$ ranges over the function space **Lag.** Somehow that space needs to be replaced by an at worst countably infinite collection of admissible Lagrangians. If that can be done by imposing some sort of physically realistic restriction on **Lag** to yield a space $\langle\mathbf{Lag}\rangle$ with cardinality $\aleph_0$, then it would be possible to build a rationalized Ultramegamultiverse starting from $V_i$ defined by

$$\langle\mathbf{UMMV_i}\rangle := \bigcup_{\mathscr{L}\in\langle\mathbf{Lag}\rangle}\langle\mathbf{MMV_i(EP(\mathscr{L}(N))}\rangle/\sim' ,$$

where $\sim'$ is the equivalence relation which glues tails of the Megamultiverses together. $\langle\mathbf{UMMV_i}\rangle$ is a four-dimensional connected manifold, with a countable basis for its topology. The cardinality of the number of branches in this space is $\aleph_0$. So it should be possible for us to populate this Ultramegamultiverse with as many as $\aleph_0$ primordial particles going into each branch of this multiverse from our original egg, which had $\aleph_0$ particles in it when t=0. This follows from the fact that $\aleph_0 \cdot \aleph_0 = \aleph_0$.

So all we need do is find some restrictions on *Lag* to make the restricted space countable. I suggest starting by requiring the elements of the restricted set of Lagrangians to be scalar densities which yield field equations which are no more that second-order in the derivatives of the field variables. Doing this will not be an easy task.

**Acknowledgements**

I would like to thank amateur astronomer Mr. Robert Roudebush for numerous discussions on the topics addressed in **section 3**.